\documentclass{article}
\pdfoutput=1
    \PassOptionsToPackage{numbers, compress}{natbib}

\usepackage[preprint]{neurips_2021}

\usepackage[utf8]{inputenc} 
\usepackage[T1]{fontenc}    
\usepackage{hyperref}       
\usepackage{url}            
\usepackage{booktabs}       
\usepackage{amsfonts}       
\usepackage{nicefrac}       
\usepackage{microtype}      
\usepackage{xcolor}         
\usepackage{float}
\usepackage{amsmath}
\usepackage{graphicx}
\usepackage{subcaption}
\usepackage{tikz}
\usetikzlibrary{bayesnet}
\usetikzlibrary{matrix}
\usepackage{nicematrix} 
\usepackage{stackengine}
\usepackage{wrapfig}
\usepackage{algorithm}
\usepackage{algpseudocode}

\DeclareMathOperator{\len}{len}

\usetikzlibrary{arrows, fit, matrix, positioning, shapes, backgrounds,
}

\definecolor{myred}{RGB}{204,102,119}
\definecolor{mygreen}{RGB}{0,158,115}
\definecolor{lightred}{RGB}{213,94,0}
\title{Parallel  Neural Local Lossless Compression}

\author{%
  Mingtian Zhang \textsuperscript{1,2}\And
  James Townsend \textsuperscript{1} \And
     Ning Kang \textsuperscript{2} \And 
     David Barber \textsuperscript{1}
  \Aff
\textsuperscript{1}AI Center, University College London, \\
 \texttt{\{m.zhang,james.townsend,d.barber\}@cs.ucl.ac.uk}\\
\textsuperscript{2}Huawei Noah’s Ark Lab\\
\texttt{kang.ning2@huawei.com}
}

\begin{document}

\maketitle

\begin{abstract}
The recently proposed Neural Local Lossless Compression (NeLLoC)~\cite{zhang2021out}, which is based on a local autoregressive model, has achieved state-of-the-art (SOTA) out-of-distribution (OOD) generalization performance in the image compression task. In addition to the encouragement of OOD generalization, the local model also allows parallel inference in the decoding stage. In this paper, we propose two parallelization schemes for local autoregressive models. We discuss the practicalities of implementing the schemes and provide experimental evidence of significant gains in compression runtime compared to the previous, non-parallel implementation. The implementations of the proposed parallelized compression methods are available at \url{https://github.com/zmtomorrow/ParallelNeLLoC}.
\end{abstract}

\section{Introduction}
Lossless compression is an important application of probabilistic modelling. Given a model $p_\theta(x)$ that  approximates an underlying distribution $p_d(x)$, a lossless compressor can be built using $p_\theta$, with compression length approximately equal to $ -\log_2p_\theta(x')$ for test data $x'$. Many models have been successfully used to compress image data, e.g. Variational Auto-Encoder (VAE) \citep{kingma2013auto} based compressors \citep{townsend2019practical,townsend2019hilloc,kingma2019bit, ruan2021improving,kingma2021variational} or flow based compressors \citep{hoogeboom2019integer,berg2020idf++,zhang2021iflow}, achieving significantly better compression rates than traditional codecs such as PNG~\citep{boutell1997png} or WebP~\citep{lian2012webp}.

Most of the recent works on model-based compressors have focused on compressing test data $x'$ that comes from the same distribution as the training data $x'\sim p_d(x)$ \citep{townsend2019practical,kingma2019bit,ruan2021improving,hoogeboom2019integer,berg2020idf++,zhang2021iflow}. However, in practice, in downstream tasks data may come from a different distribution $x'\sim p_o\neq p_d$, and we would like to design probabilistic models with good out-of-distribution (OOD) generalization ability, whilst still achieving near state-of-the-art results on in-distribution test data. 

Recent work \citep{zhang2021out} shows that it is possible for image models to generalize well to OOD data based on the following hypothesis \citep{schirrmeister2020understanding, havtorn2021hierarchical,zhang2021out,zhang2022out}: \emph{local features (like smoothness, edges, corners etc.) are shared by different (natural) image distributions and dominate the log-likelihood,}
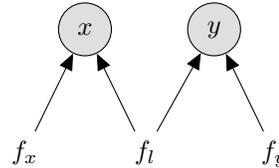
\begin{wrapfigure}{r}{0.5\textwidth}
\vspace{-0.1cm}
\centering
    \begin{tikzpicture}[square/.style={regular polygon,regular polygon sides=4}]
         \node[obs] (x1) {$x$};
          \node[obs, right=of x1] (x2) {$y$};
\node[ below=of x1,xshift=0.8cm] (z1) {$f_{l}$};

           \node[ below=of x1,xshift=-0.8cm] (z2) {$f_x$};
           \node[ below=of x2,xshift=0.8cm] (z3)
           {$f_y$};
         \edge {z1} {x1};
          \edge {z1} {x2};
          \edge {z2} {x1};
          \edge {z3} {x2};
    \end{tikzpicture}
        \label{fig:my_label}
        \captionof{figure}{Graphical visualization of the hypothesis. We use $x$ and $y$ to represent two different image distributions $p(x), p(y)$; $f_l$ to indicate the shared local features and $f_x, f_y$ are the non-local features that are specific to  $p(x)$ and $p(y)$ respectively. \label{fig:hypo}}
        \vspace{-0.4cm}
\end{wrapfigure}
\emph{whereas non-local features (like semantics) are not shared.} 
Figure \ref{fig:hypo} gives a graphical illustration of the hypothesis. This hypothesis suggests that learning the non-local features that are specific to one distribution may hurt the OOD 
 generalization performance, which has also been empirically verified in \citep{zhang2021out}.
Therefore, to encourage OOD generalization, a local model can be used to build the lossless compressor. Under this intuition, paper \citep{zhang2021out} proposed the Neural Local Lossless Compressor (NeLLoC), which uses an autoregressive model with local dependency to ensure it can only learn local features. 

In addition to the OOD generalization encouragement, the local autoregressive models also have other fruitful properties for compression applications such as extremely small model size and the ability to compress images with arbitrary sizes, we refer readers to \cite{zhang2021out} for a detailed discussion. However, the decompression time of the original NeLLoC scales with  $O(D^2)$ for a $D\times D$ image, which is a computation bottleneck for large images. In this paper, we propose a parallelization scheme that can improve the complexity from $O(D^2)$ to $O(D)$ on a machine without changing the compression rate.
In the next section, we  give a brief introduction to lossless compression with autoregressive models.

\section{Lossless Compression with Autoregressive Models}
Assuming each image $\mathbf{x}$ has  $D\times D$  byte valued pixels $x_{ij}\in \{0,\cdots,255\}$. A \emph{full autoregressive model} can be written as 
\begin{align}
p_f(\mathbf{x})=\prod_{ij}p(x_{ij}|x_{[1:i-1,1:D]},x_{[i,1:j-1]}),
\end{align}
where $x_{[1:i-1,1:D]}=\emptyset$ when $i=1$ and $x_{[i,1:j-1]}=\emptyset$ when $j=1$.
In practice, this full autoregressive model can be implemented by various structures, such as PixelRNN~\citep{van2016pixel}, PixelCNN~\citep{van2016pixel,salimans2017pixelcnn++} or Transformers~\citep{child2019generating}. The model is then trained by minimizing  the KL divergence between the data distribution $p_d$ and the model $p_f$, which can be further approximated by a Monte-Carlo approximation with finite training data samples $\{\mathbf{x}^{1},\cdots, \mathbf{x}^{N}\}\sim p_d(\mathbf{x})$
\begin{align}
\mathrm{KL}(p_d(\mathbf{x})||p_f(\mathbf{x}))&=-\int p_d(\mathbf{x}) \log p_f(\mathbf{x}) dx-\mathrm{H}(p_d)\approx -\frac{1}{N} \sum_{n=1}^N \log p_f(\mathbf{x}^n) +const.,
\end{align}
where the entropy of the data distribution $\mathrm{H}(p_d)$ is a constant. Once we have learned the model $p_f(\mathbf{x})$,  a stream coder is introduced to build a bijection between a given image $\mathbf{x}'$ and a binary  string $s$ with length $\len(s)\approx -\log_2 p_f(\mathbf{x}')$. When  $p_f\rightarrow p_d$, the length $-\log_2 p_f(\mathbf{x}')$ approaches the optimal compression length under Shannon's source coding theorem~\cite{shannon2001mathematical}. Popular stream coders include 
 Arithmetic Coding (AC) \citep{witten1987arithmetic} and Asymmetric Numeral System (ANS) \citep{duda2013asymmetric}, we refer the reader to \citep{mackay2003information} and \citep{townsend2020tutorial} for a detailed introduction of AC and ANS coders respectively.

Note that, for a full autoregressive model,  the decoding of $i$th pixel of image $\mathbf{x}'$ requires the distribution of $x_i$, which requires firstly decoding all the previous pixels $x'_{1:i-1}$ to infer  $p(x_i|x'_1,\cdots,x'_{i-1})$.
Therefore, the sequential inference mechanism in the decoding stage scales with $\mathcal{O}(D^2)$.

\subsection{Local Autoregressive Models}
We define a \emph{local autoregressive model} $p_l(x)$ with \emph{dependency horizon} $h$\footnote{The dependency horizon $h$ can also be interpreted as the \emph{Chebyshev distance}~\cite{cantrell2000modern} between pixels.}, to be an autoregressive model where each pixel depends only on previous pixels in a local region specified by $h$. To be precise, local model $p_l(x)$ can be written as
\begin{align}
  p_l(\mathbf{x})=\prod_{ij}p(x_{ij}|x_{[i-h:i-1,j-h:j+h]},x_{[i,j-h:j-1]}), \label{eq:local:model}
\end{align}
with zero-padding used in cases where $i$ or $j$ are smaller than $h$. Figure \ref{fig:full:local}b shows the dependency relationship between pixels in a local autoregressive model, compared to a full autoregressive model (Figure \ref{fig:full:local}a), where each pixel depends on all previous pixels. 

\begin{figure}[h]
    \centering
     \begin{subfigure}[b]{0.45\textwidth}
    \centering
    \begin{tikzpicture}[scale=0.9, every node/.style={scale=0.9}]
    \draw[xstep=.8cm,ystep=.8,color=gray] (0,0) grid (4,4);
    \matrix (M) [matrix of nodes,
    inner sep=0pt,
    anchor=south west,
    nodes={inner sep=0pt,text width=.8cm,align=center,minimum height=.8cm},
    ]{
    $x_{11}$ & $x_{12}$ & $x_{13}$ &  $x_{14}$ & $x_{15}$\\
    $x_{21}$ & $x_{22}$ & $x_{23}$ &  $x_{24}$ & $x_{25}$\\
    $x_{31}$ & $x_{32}$ & $x_{33}$ &  $x_{34}$ & $x_{35}$\\
    $x_{41}$ & $x_{42}$ & $x_{43}$ &  $x_{44}$ & $x_{45}$\\
    $x_{51}$ & $x_{52}$ & $x_{53}$ &  $x_{54}$ & $x_{55}$\\
    };
        \scoped[on background layer]
        {
        \fill[mygreen] (0,2.4) rectangle (4,4);
        \fill[mygreen] (0,1.6) rectangle (1.6,2.4);
        \fill[myred] (1.6,1.6) rectangle (2.4,2.4);
        }
    \end{tikzpicture}
 
    \caption{Full autoregressive model\label{fig:full:locala}}
    \end{subfigure}
     \begin{subfigure}[b]{0.45\textwidth}
    \centering
    \begin{tikzpicture}[scale=0.9, every node/.style={scale=0.9}]
    \draw[xstep=.8cm,ystep=.8,color=gray] (0,0) grid (4,4);
    \matrix (M) [matrix of nodes,
    inner sep=0pt,
    anchor=south west,
    nodes={inner sep=0pt,text width=.8cm,align=center,minimum height=.8cm},
    ]{
    $x_{11}$ & $x_{12}$ & $x_{13}$ &  $x_{14}$ & $x_{15}$\\
    $x_{21}$ & $x_{22}$ & $x_{23}$ &  $x_{24}$ & $x_{25}$\\
    $x_{31}$ & $x_{32}$ & $x_{33}$ &  $x_{34}$ & $x_{35}$\\
    $x_{41}$ & $x_{42}$ & $x_{43}$ &  $x_{44}$ & $x_{45}$\\
    $x_{51}$ & $x_{52}$ & $x_{53}$ &  $x_{54}$ & $x_{55}$\\
    };
        \scoped[on background layer]
        {
        \fill[mygreen] (0.8,2.4) rectangle (3.2,3.2);
        \fill[mygreen] (0.8,1.6) rectangle (1.6,2.4);
        \fill[myred] (1.6,1.6) rectangle (2.4,2.4);
        }
\draw [thick, red] (0.8,0.8) -- (0.8,3.2);
\draw [thick, red] (0.8,3.2) -- (3.2,3.2);
\draw [thick, red] (3.2,3.2) -- (3.2,0.8);
\draw [thick, red] (3.2,0.8) -- (0.8,0.8);
    \end{tikzpicture}
    \caption{Local autoregressive model with $h=1$\label{fig:full:localb}}
    \end{subfigure}
    \caption{Comparison of full  and local autoregressive models. In a full autoregressive model (a), the red pixel depends on all the previous pixels (in green), whereas in a local autoregressive model (b), the red pixel only depends on the previous pixels in a local region (in green).\label{fig:full:local}}
    \end{figure}

The local autoregressive model can be efficiently implemented by a simple modification of the PixelCNN structure~\citep{van2016pixel,zhang2021out}. Specifically, given a dependency horizon $h$, the local PixelCNN can be constructed by letting the first masked convolutional layer have kernel height $(h+1)$ and width $2h+1$ and letting the subsequent layers be $1\times 1$ convolutions. In this case, the dependency horizon just depends on the kernel size of the first masked convolutional kernel. 

The compression implementation provided in \citep{zhang2021out} uses a sequential inference mechanism that is the same as in the full autoregressive model, with time complexity in $\mathcal{O}(D^2)$. However, for a local autoregressive model, there exist pixels that are conditional independent given the previous observed (decoded) pixels, allowing parallel inference in the decoding stage. In the following sections, we introduce an exact inference mechanism with runtime which scales with $\mathcal{O}(D)$ on a parallel machine and demonstrate a significant speedup experimentally.

\section{Parallel Decoding with Local Autoregressive Models}
In contrast to full autoregressive models, where pixels must be decoded sequentially, there exist pixels that can be independently decoded in a local autoregressive model. Figure \ref{fig:topo} gives the topological order of parallel decoding, for a  $5\times 5$ image, using a local autoregressive model with $h=1$. The number in each pixel indicates the time at which the pixel can be decoded. For example, the two red pixels  marked with time $6$ can be decoded in parallel since they are  independent under the model.

\begin{wrapfigure}{r}{0.35\textwidth}
    \vspace{-0.4cm}
    \centering
    \begin{tikzpicture}[scale=0.9, every node/.style={scale=0.9}]
    \draw[xstep=.8cm,ystep=.8cm,color=gray] (0,0) grid (4,4);
    \matrix (M) [matrix of nodes,
    inner sep=0pt,
    anchor=south west,
    nodes={inner sep=0pt,text width=0.8cm,align=center,minimum height=.8cm},
    ]{
    1 & 2 & 3 & 4 & 5\\
    3 & 4 & 5 & 6 & 7\\
    5 & 6 & 7 & 8 & 9\\
    7 & 8 & 9 & 10 & 11 \\
    9 & 10 & 11 & 12 & 13 \\
    };
    \scoped[on background layer]
        {
        \fill[mygreen] (1.6,2.4) rectangle (2.4,4);
        \fill[mygreen] (2.4,3.2) rectangle (4,4);
        \fill[mygreen] (0,2.4) rectangle (3.2,3.2);
        \fill[mygreen] (0,2.4) rectangle (3.2,3.2);
        \fill[mygreen] (0,1.6) rectangle (0.8,2.4);
        \fill[myred] (0.8,1.6) rectangle (1.6,2.4);
        \fill[myred] (2.4,2.4) rectangle (3.2,3.2);
        \draw [thick, red] (0,0.8) -- (0,3.2);
        \draw [thick, red] (2.4,0.8) -- (2.4,3.2);
        \draw [thick, red] (0,0.8) -- (2.4,0.8);
        \draw [thick, red] (0,3.2) -- (2.4,3.2);
        \draw [thick, red] (1.6,1.6) -- (1.6,4);
        \draw [thick, red] (1.6,1.6) -- (4,1.6);
        \draw [thick, red] (4,1.6) -- (4,4);
        \draw [thick, red] (1.6,4) -- (4,4);
        }
    \end{tikzpicture}
     \caption{Topological decoding order. Pixels with the same number are parallel decoded.\label{fig:topo}}
    \vspace{-0.5cm}
    \end{wrapfigure}
In general, for an image with size $D\times D$, on a machine with $\lfloor \frac{D+h}{h+1}\rfloor$ parallel processing units, the total decoding time $T=D+(D-1)\times (h+1)$. Since $h$ is a small constant $h\ll D$, the decoding time scales with $\mathcal{O}(D)$, which is a significant improvement over the $\mathcal{O}(D^2)$ of full autoregressive models. In the example in Figure \ref{fig:topo}, for a $5\times 5$ image with dependency length $h=1$, the decoding time is $T=13$ whereas in a full autoregressive model, $T=25$. In the next section, we discuss how to implement the parallelization scheme in practice.

We observe that for fixed \(h\), the positions of pixels decoded at each time step do not change. We can thus pre-compute the topological ordering and save the locations the pixels computed at each time step. At each time step in the decoding stage, we just load the saved positions of the independent pixels to be decoded and collect the image patches on which they depend into a batch. For example, for the pixel $x_{33}$ in Figure \ref{fig:full:localb}, the relevant patch is a $3\times 3$ square marked in red, the redundant pixels $\{x_{33},x_{34},x_{42},x_{43},x_{44}\}$ will be masked out in the local autoregressive model. One can also take a rectangle patch that contains $\{x_{22},x_{23},x_{24},x_{32},x_{33},x_{34}\}$ to reduce the computation and $\{x_{33},x_{34}\}$ will be masked out in this case. We pad the patches with $0$ for the pixels  near the boundary, to make sure all the patches have the same size. The parallel decoding procedure is summarized in Algorithm \ref{algo:pdecoding}.

 \begin{algorithm}
\caption{Parallel Decoding Procedure of Local Autoregressive Models \label{algo:pdecoding}}
  \begin{algorithmic}[1]
    \For{t = 1 to T}
    \State Load the positions  of independent pixels to be decoded at time $t$.
        \State Gather the relevant patches based on the loaded positions to form a batch.
        \State In parallel, compute the predictive distributions for those pixels using the batch.
        \State Decode the pixel values using the predictive distributions.
      \EndFor 
  \end{algorithmic}
\end{algorithm}
  \vspace{-0.2cm}




\subsection{Sheared Local Autoregressive Model}
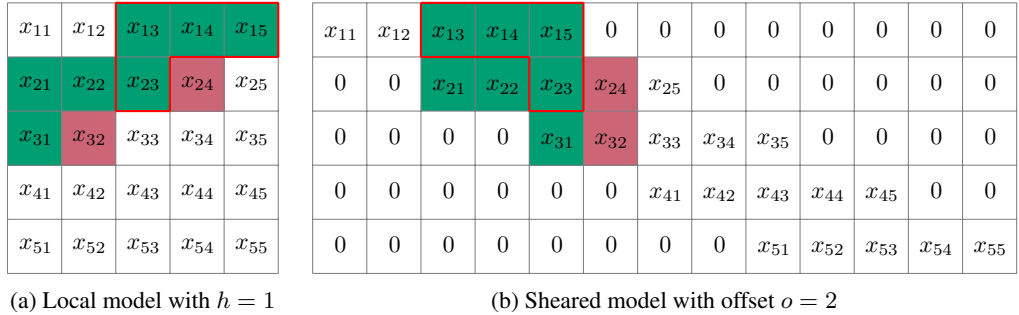
\begin{figure}[H]
\vspace{-0.1cm}
    \centering
     \begin{subfigure}[b]{0.29\textwidth}
    \centering
    \begin{tikzpicture}[scale=0.9, every node/.style={scale=0.9}]
    \draw[xstep=.8cm,ystep=.8,color=gray] (0,0) grid (4,4);
    \matrix (M) [matrix of nodes,
    inner sep=0pt,
    anchor=south west,
    nodes={inner sep=0pt,text width=.8cm,align=center,minimum height=.8cm},
    ]{
    $x_{11}$ & $x_{12}$ & $x_{13}$ &  $x_{14}$ & $x_{15}$\\
    $x_{21}$ & $x_{22}$ & $x_{23}$ &  $x_{24}$ & $x_{25}$\\
    $x_{31}$ & $x_{32}$ & $x_{33}$ &  $x_{34}$ & $x_{35}$\\
    $x_{41}$ & $x_{42}$ & $x_{43}$ &  $x_{44}$ & $x_{45}$\\
    $x_{51}$ & $x_{52}$ & $x_{53}$ &  $x_{54}$ & $x_{55}$\\
    };
        \scoped[on background layer]
        {
        \fill[mygreen] (1.6,2.4) rectangle (2.4,4);
        \fill[mygreen] (2.4,3.2) rectangle (4,4);
        \fill[mygreen] (0,2.4) rectangle (3.2,3.2);
        \fill[mygreen] (0,2.4) rectangle (3.2,3.2);
        \fill[mygreen] (0,1.6) rectangle (0.8,2.4);
        \fill[myred] (0.8,1.6) rectangle (1.6,2.4);
        \fill[myred] (2.4,2.4) rectangle (3.2,3.2);
        }
        \draw [thick, red] (1.6,2.4) -- (2.4,2.4);
        \draw [thick, red] (2.4,2.4) -- (2.4,3.2);
        \draw [thick, red] (1.6,2.4) -- (1.6,4.0);
        \draw [thick, red] (2.4,3.2) -- (4.0,3.2);
        \draw [thick, red] (4.0,3.2) -- (4,4);
        \draw [thick, red] (1.6,4.0) -- (4,4);
    \end{tikzpicture}
    \caption{Local model with $h=1$\label{fig:local}}
    \end{subfigure}
     \begin{subfigure}[b]{0.69\textwidth}
    \centering
    \begin{tikzpicture}[scale=0.9, every node/.style={scale=0.9}]
    \draw[xstep=.8cm,ystep=.8cm,color=gray] (0,0) grid (10.4,4.0);
    \matrix (M) [matrix of nodes,
    inner sep=0pt,
    anchor=south west,
    nodes={inner sep=0pt,text width=.8cm,align=center,minimum height=.73cm},
    ]{
    $x_{11}$ & $x_{12}$ & $x_{13}$ &  $x_{14}$ & $x_{15}$ & $0$ & $0$&$0$ & $0$&$0$&$0$&$0$&$0$\\
    $0$&$0$&$x_{21}$ & $x_{22}$ & $x_{23}$ &  $x_{24}$ & $x_{25}$ & $0$ & $0$&$0$&$0$&$0$&$0$\\
    $0$&$0$&$0$&$0$&$x_{31}$ & $x_{32}$ & $x_{33}$ &  $x_{34}$ & $x_{35}$& $0$&$0$&$0$&$0$\\
    $0$&$0$&$0$&$0$&$0$&$0$&$x_{41}$ & $x_{42}$ & $x_{43}$ &  $x_{44}$ & $x_{45}$&$0$&$0$\\
    $0$&$0$&$0$&$0$&$0$&$0$&$0$&$0$&$x_{51}$ & $x_{52}$ & $x_{53}$ &  $x_{54}$ & $x_{55}$\\
    };
    \scoped[on background layer]
        {
        \fill[mygreen] (1.6,2.4) rectangle (4.0,4.0);
        \fill[mygreen] (3.2,1.6) rectangle (4.0,2.4);
        \fill[myred] (4.0,1.6) rectangle (4.8,3.2);
        }
        \draw [thick, red] (1.6,4.0) -- (4.0,4.0);
        \draw [thick, red] (4.0,2.4) -- (4.0,4.0);
        \draw [thick, red] (4.0,2.4) -- (3.2,2.4);
        \draw [thick, red] (3.2,3.2) -- (3.2,2.4);
        \draw [thick, red] (3.2,3.2) -- (1.6,3.2);
        \draw [thick, red] (1.6,4.0) -- (1.6,3.2);
    \end{tikzpicture}
    \caption{Sheared model with offset $o=2$}
    \end{subfigure}
    \caption{Pixel dependency after the shear operation. The red pixels in the same column in the sheared image (b) are conditionally independent given the green pixels, and are aligned in memory. \label{fig:shear}}
    \vspace{-0.3cm}
    \end{figure}

We notice that in Algorithm \ref{algo:pdecoding}, the conditionally independent pixels in each step are located in nonadjacent positions in the images. 
For example, the dependent areas (green) of the two red pixels in Figure \ref{fig:topo} are not aligned in memory. This requires extra indexing time when reading/writing their values. 
To alleviate the potential speed limitation,  we propose to transform the model such that the conditionally independent pixels are aligned. Specifically, for a local autoregressive model with dependency horizon $h$, we shear the model, and image, with offset $o=h+1$.  Figure \ref{fig:shear} shows an example where the local autoregressive model with $h=1$ (Figure \ref{fig:local}) is sheared with  offset $o=2$.  We observe that in the sheared model, the conditionally independent pixels are aligned in memory, allowing significantly faster parallel reading/writing of those pixels. The sheared model has length $L=D+(D-1)\times o$ for a $D\times D$ images, which is equal to the decoding steps T in pNeLLoC since $o=h+1$. Therefore, the inference time also scales with $O(D)$ on parallel processing units.

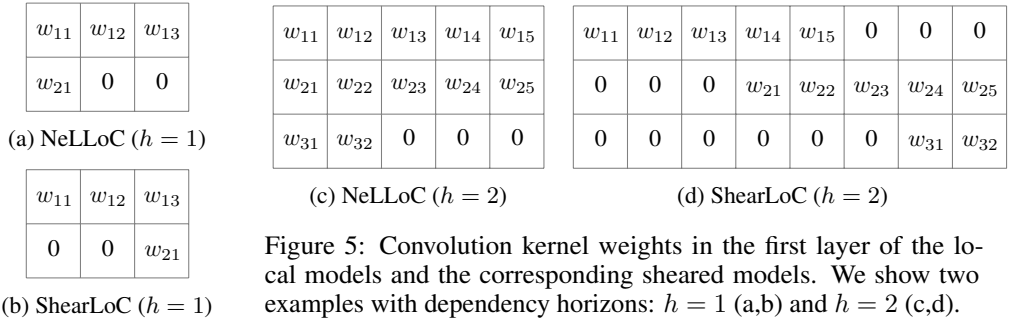
\begin{figure}[H]
\vspace{-0.3cm}
\begin{tabular}{c}
\begin{subfigure}[t]{0.2\textwidth}
    \centering
    \begin{tikzpicture}[scale=0.9, every node/.style={scale=0.9}]
    \draw[xstep=.8cm,ystep=.8cm,color=gray] (0,0) grid (2.4,1.6);
    \matrix (M) [matrix of nodes,
    inner sep=0pt,
    anchor=south west,
    nodes={inner sep=0pt,text width=.8cm,align=center,minimum height=.73cm},
    ]{
    $w_{11}$ & $w_{12}$ & $w_{13}$ \\
    $w_{21}$ &  0 &  0 \\
    };
    \end{tikzpicture}
    \caption{NeLLoC ($h=1$)}
    \vspace{0.2cm}
    \end{subfigure}\\
     \begin{subfigure}[c]{0.2\textwidth}
         \centering
         \begin{tikzpicture}[scale=0.9, every node/.style={scale=0.9}]
    \draw[xstep=.8cm,ystep=.8cm,color=gray] (0,0) grid (2.4,1.6);
    \matrix (M) [matrix of nodes,
    inner sep=0pt,
    anchor=south west,
    nodes={inner sep=0pt,text width=.8cm,align=center,minimum height=.73cm},
    ]{
    $w_{11}$ & $w_{12}$ & $w_{13}$\\
    0&0& $w_{21}$\\
    };
    \end{tikzpicture}
    \caption{ShearLoC ($h=1$)}
    \end{subfigure}
    \end{tabular}
    \begin{tabular}{c}
     \begin{subfigure}[t]{0.3\textwidth}
         \vspace{0.01cm}
    \centering
    \begin{tikzpicture}[scale=0.9, every node/.style={scale=0.9}]
    \draw[xstep=.8cm,ystep=.8,color=gray] (0,0) grid (4.,2.4);
    \matrix (M) [matrix of nodes,
    inner sep=0pt,
    anchor=south west,
    nodes={inner sep=0pt,text width=.8cm,align=center,minimum height=.76cm},
    ]{
    $w_{11}$ & $w_{12}$ & $w_{13}$ &  $w_{14}$ & $w_{15}$\\
    $w_{21}$ & $w_{22}$ & $w_{23}$ &  $w_{24}$ & $w_{25}$\\
    $w_{31}$ & $w_{32}$ & 0 &  0 & 0\\
    };
    \end{tikzpicture}
    \caption{NeLLoC ($h=2$)}
    \end{subfigure}
     \begin{subfigure}[t]{0.4\textwidth}
         \vspace{0.01cm}
         \centering
         \begin{tikzpicture}[scale=0.9, every node/.style={scale=0.9}]
    \draw[xstep=.8cm,ystep=.8,color=gray] (0,0) grid (6.4,2.4);
    \matrix (M) [matrix of nodes,
    inner sep=0pt,
    anchor=south west,
    nodes={inner sep=0pt,text width=.8cm,align=center,minimum height=.73cm},
    ]{
    $w_{11}$ & $w_{12}$ & $w_{13}$ &  $w_{14}$ & $w_{15}$ & 0 & 0 & 0\\
    0&0&0& $w_{21}$ & $w_{22}$ & $w_{23}$ &  $w_{24}$ & $w_{25}$\\
      0&0&0&0&0&0&$w_{31}$ & $w_{32}$ \\
    };
    \end{tikzpicture}
    \caption{ShearLoC ($h=2$)}
    \end{subfigure}\\
    \begin{minipage}{.68\textwidth}
    \vspace{0.3cm}
\caption{Convolution kernel weights in the first layer of the local  models and the corresponding sheared models. We show two examples with  dependency horizons: $h=1$ (a,b) and $h=2$ (c,d).\label{fig:shear:cnn}}
\end{minipage}
\end{tabular}
\vspace{-0.4cm}
\end{figure}

As discussed in Section 2, the pixel dependency structure of the local autoregressive model only depends on the first convolution kernel. Therefore, to shear the model, we only need to shear the first convolution kernel. Figure \ref{fig:shear:cnn} visualizes the sheared convolutional kernel for two local autoregressive models with $h=1$ and $h=2$. After shearing the model, we also need to shear the images to conduct compression and decompression, see Algorithm \ref{algo:sdecoding} for a summary of the decoding procedure. We refer to compression with the sheared  model as \textbf{Shear}ed \textbf{Lo}cal Lossless \textbf{C}ompression (ShearLoC). 

 \begin{algorithm}[t]
\caption{Parallel Decoding Procedure of Sheared Local Autoregressive Models \label{algo:sdecoding}}
  \begin{algorithmic}[1]
  \State Shear the image based on the dependency horizon.
    \For{t = 1 to T}
        \State In parallel, compute the predictive distributions for those pixels in each column.
        \State Decode the pixel values using the predictive distributions.
      \EndFor 
\State Undo the shear operation for the decoded image.
  \end{algorithmic}
\end{algorithm}

\section{Demonstrations}
We implement the compression model using PyTorch~\citep{NEURIPS2019_9015} with ANS. For the local autoregressive model, we use the same model architecture as that used in \citep{zhang2021out}:  a local PixelCNN  with  horizon $h=3$, followed by ResNet Blocks with $1\times1$ convolution, see  Appendix A.3 in \citep{zhang2021out} for details. 
We also use the three pre-trained (on CIFAR10) models (with 0, 1 and 3 ResNet blocks) provided by \citep{zhang2021out} for all the experiments.  The model with 3 ResNet blocks was shown by \citep{zhang2021out} to achieve SOTA OOD compression rate, see Table 6 of \citep{zhang2021out} for a detailed comparison with other compression methods.
\begin{wraptable}{r}{5.5cm}
\vspace{-.3cm}
\caption{Test BPD on CIFAR10\label{tab:BPD}}
    \begin{tabular}{lllll}
        \toprule
         Res. Num. & 0   &  1  & 3 \\
         Size (MB) & 0.49& 1.34 &2.75\\
          \midrule
         BPD & 3.38 & 3.28 & 3.25 \\ 
         \bottomrule
    \end{tabular}
    \vspace{-0.2cm}
\end{wraptable}
 For reference, we report the model sizes and the test BPD on CIFAR10 for these three models in Table \ref{tab:BPD}.
During encoding, since all pixels are observed, the statistics of all the pixels can be computed in parallel. However, we have found that on CPU the computations may not be deterministic when using different batch sizes during the encoding and decoding. Therefore, we instead use the an identical inference procedure in both the encoding and decoding stages. Other  details can be found in the provided repository. All the experiments are conducted on a MacBook Air (2020) with M1 chip and 8GB memory, the results are averaged over 10 images from the ImageNet~\cite{deng2009imagenet} dataset.

We compare parallel NeLLoC (pNeLLoC) and ShearLoC with the original sequential NeLLoC (sNeLLoC) implementation. Since all algorithms use the same underlying pre-trained model, they have the same compression BPD.  Table \ref{tab:model:sizes} shows the decompression time comparison using three models on images with side length 32. We find pNeLLoC  is 2x faster than sNeLLoC  with the 0 ResNet block model, and the improvement increases for larger models. Compared to pNeLLoC, ShearLoC achieves a further speed improvement, with a more significant advantage in larger models.

We  also compare the decompression time on square images with increased side lengths: $[32, 64, 128, 1024]$\footnote{The 1024\(\times\)1024 images  are provided in the repository, the corresponding result is averaged over 3 images.}. Table \ref{tab:image:size} shows the improvement percentage from using pNeLLoC grows when we increase the size of the test images. This is consistent with the theoretical argument that the proposed parallelization scheme improves the computation complexity from $O(D^2)\rightarrow O(D)$ on parallel units. Similarly, additional improvements can be achieved when using ShearLoC, which shares the same complexity with pNeLLoC but has more efficient memory access.
\begin{table}[h]
\vspace{-0.2cm}
        \centering
\captionsetup{width=.8\textwidth}
     \caption{Decoding time (s) comparisons. We show the improvement  of pNeLLoC (in green) and ShearLoC (in red) comparing to using sNeLLoC.\label{tab:time}}
    \begin{subtable}[h]{\textwidth}
        \centering
        \caption{Different model sizes, the image has size $32\times 32$.\label{tab:model:sizes}}
        \begin{tabular}{lllll}
    \toprule
    Res. Num. & 0   &  1  & 3 \\
    \midrule
    BPD & 3.39 & 3.32 & 3.29\\
    \midrule
   sNeLLoC &  0.460 (-) & 0.578 (-) & 0.774 (-) \\ 
   pNeLLoC  &  0.223 (\textcolor{mygreen}{2.06x}) & 0.277 (\textcolor{mygreen}{2.09x}) & 0.335 (\textcolor{mygreen}{2.31x}) \\
   ShearLoC & 0.218 (\textcolor{myred}{2.11x}) & 0.222 (\textcolor{myred}{2.60x})&0.245 (\textcolor{myred}{3.16x})\\
    \bottomrule
  \end{tabular}
    \end{subtable}
    
     \begin{subtable}[h]{\textwidth}
        \centering
        \caption{Different image sizes, the model has 0 ResNet blocks.\label{tab:image:size}}
        \begin{tabular}{llllll}
    \toprule
     Side len. & 32   &  64  & 128 & 1024\\
     \midrule
     BPD & 3.39 & 3.05 &2.93 & 2.22\\
    \midrule
   sNeLLoC &  0.460 (-) & 1.879 (-)&  7.574 (-) & 475.9 (-)\\ 
   pNeLLoC &  0.223  (\textcolor{mygreen}{2.06x}) & 0.757 (\textcolor{mygreen}{2.48x}) & 2.217  (\textcolor{mygreen}{3.42x}) &100.0 (\textcolor{mygreen}{4.58x})\\
      ShearLoC & 0.218 (\textcolor{myred}{2.11x}) & 0.612 (\textcolor{myred}{3.07x}) & 1.683 (\textcolor{myred}{4.60x})& 73.00 (\textcolor{myred}{6.52x})\\

    \bottomrule
  \end{tabular}
     \end{subtable}
     \vspace{-0.2cm}
\end{table}

\section{Discussion}
Several methods have been proposed to improve the sampling runtime in autoregressive models. For example, \citep{reed2017parallel,razavi2019generating} explore the multi-scale structure in the image domain, and design models that allow parallel generation of pixels in higher resolution samples conditioned on low resolution samples. In contrast to previous works, the parallel method proposed in this paper is specially designed for local autoregressive models, with the flexibility to handle images of arbitrary size. Local autoregressive models can also be combined with latent variable models to generate semantically-coherent images~\citep{gulrajani2016pixelvae,zhang2022improving}. In this case, the proposed parallelization schemes can be also used to improve the sampling efficiency, which we leave to future work.

\clearpage 
\newpage
\bibliographystyle{abbrvnat}
\bibliography{main.bib}

\appendix



\end{document}